\documentclass[twocolumn,showpacs,preprintnumbers,amsmath,amssymb]{revtex4}
\usepackage{graphicx}
\usepackage{dcolumn}
\usepackage{bm}

\begin{document}

\title{Systematic {\it ab initio} study of the magnetic and electronic properties 
of all 3$d$ transition metal linear and zigzag nanowires}
\author{J. C. Tung and G. Y. Guo\footnote{E-mail: gyguo@phys.ntu.edu.tw} }
\affiliation{Department of Physics and Center for Theoretical Sciences, National Taiwan University, Taipei 
106, Taiwan}

\date{\today}

\begin{abstract}
The magnetic and electronic properties of both linear and zigzag atomic chains of all 3$d$ transition 
metals have been calculated within density
functional theory with the generalized gradient approximation.
The underlying atomic structures were determined theoretically. 
It is found that all the zigzag chains except the nonmagnetic (NM) Ni
and antiferromagnetic (AF) Fe chains which form a twisted two-legger ladder, 
look like a corner-sharing triangle ribbon, and have a lower 
total energy than the corresponding linear chains.
All the 3$d$ transition metals in both linear and zigzag structures
have a stable or metastable ferromagnetic (FM) state.
Furthermore, in the V, Cr, Mn, Fe, Co linear chains
and Cr, Mn, Fe, Co, Ni zigzag chains, a stable or metastable AF state also exists.
In the Sc, Ti, Fe, Co, Ni linear structures, the FM state is the ground state whilst
in the V, Cr and Mn linear chains, the AF state is the ground state.
The electronic spin-polarization at the Fermi level in the FM Sc, V, Mn, Fe,
Co and Ni linear chains is close to 90\% or above, suggesting that these nanostructures
may have applications in spin-transport devices. Interestingly,
the V, Cr, Mn, and Fe linear chains show a giant magneto-lattice expansion
of up to 54 \%. In the zigzag structure, the AF state
is more stable than the FM state only in the Cr chain.   
Both the electronic magnetocrystalline anisotropy 
and magnetic dipolar (shape) anisotropy energies 
are calculated. It is found that the shape anisotropy energy
may be comparable to the electronic one and always prefers the axial magnetization
in both the linear and zigzag structures. 
In the zigzag chains, there is also a pronounced shape anisotropy in the
plane perpendicular to the chain axis.
Nonetheless, in the FM Ti, Mn, Co and
AF Cr, Mn, Fe linear chains, the electronic anisotropy is perpendicular,
and it is so large in the FM Ti and Co as well as AF Cr, Mn and Fe linear chains that the
easy magnetization axis is perpendicular.
In the AF Cr and FM Ni zigzag structures, the easy magnetization direction
is also perpendicular to the chain axis but in the ribbon plane.
Remarkably, the axial magnetic anisotropy in the FM Ni linear chain is
gigantic, being $\sim$ 12 meV/atom, suggesting that Ni nanowires may have
applications in ultrahigh density magnetic memories and hard disks.
Interestingly, there is a spin-reorientation transition in the FM Fe and Co linear
chains when the chains are compressed or elongated.
Large orbital magnetic moment is found in the FM Fe, Co and Ni linear chains.
Finally, the band structure and density of states of the nanowires
have also been calculated to identify the electronic origin
of the magnetocrystalline anisotropy and orbital magnetic moment.
\end{abstract}


\maketitle

\section{Introduction}
Magnetism at the nanometer scale has been 
a very active research area in recent years~\cite{hei00,pie00,Hern,Hern2},
because of its novel fundamental physics and exciting potential applications.
Theoretically, 
a great deal of research has been done on both finite and infinite chains 
of atoms. In particular, calculations for isotropic Heisenberg model
with finite-range exchange interactions show that a one dimensional (1D)
chain cannot maintain ferromagnetism at any finite temperatures.~\cite{Mermin}
Nonetheless, this discouraging conclusion has to be revised when
a magnetic anisotropy is present, as in, e.g., quasi-1D crystals. 
Experimentally, modern methods to prepare nanostructured systems 
have made it possible to investigate the influence
of dimensionality on the magnetic properties. A fundamental idea is to 
exploit the geometrical restriction imposed by an array of parallel steps on a 
vicinal surface along which the deposited material can nucleate.
For example, Gambardella, {\it et al.}\cite{Gambardella2}, recently 
succeeded in preparing a high
density of parallel atomic chains along steps by growing Co on a high-purity
Pt (997) vicinal surface and also observed 1D magnetism in a narrow temperature 
range of 10$\sim$20 K. 
Structurally stable nanowires can also be grown 
inside tubular structures, such as the Ag nanowires of micrometer
lengths grown inside self-assembled organic (calix[4]hydroquinone)
nanotubes\cite{Suh}. Short suspended nanowires have been produced by driving
the tip of scanning tunneling microscope into contact with a metallic surface
and subsequent retraction, leading to the extrusion of a limited number of
atoms from either tip or substrate\cite{Rubio}. 
Monostrand nanowires of Co and Pd have also been prepared in 
mechanical break junctions, and full spin-polarized conductance 
was observed\cite{Rodrigues}. 

The monoatomic chains, being an ultimate 1D structure, are a testing ground 
for the theories and concepts developed earlier for three-dimensional (3D) 
systems. Furthermore, the 1D characters of nanowires can cause several 
new physical phenomena to appear. It is of fundamental importance to 
understand the atomic structure in a truly 1D nanowire and how the 
magnetic and electronic properties change in the lower dimensionality. 
Therefore, theoretical calculations at either semi-empirical tight-binding or 
{\it ab initio} density functional theory level for many infinite/finite chains,
e.g., linear chains of Co\cite{Dallmeyer, Jisang, Matej, Lazarovits, Ederer}, 
Fe\cite{Ederer,Spisak2}, Ni, Pd\cite{Spisak}, Pt, Cu\cite{Dallmeyer}, 
Ag\cite{Ribeiro, Nautiyal2}, and Au\cite{Bahn, Delin, Ribeiro, Skorodumova, Maria}, 
as well as zigzag chains of Ti\cite{li05}, Fe\cite{Spisak2}, 
and Au\cite{Skorodumova}, have been reported. 
Early studies of infinite linear chains of 
Au \cite{Portal1, Portal2, Maria, Skorodumova}, Al \cite{Sen}, Cu\cite{Nautiyal2}, 
Ca, Pd\cite{Delin}, and K\cite{Delin} have shown a wide variety of stable and metastable 
structures. Recently, the magnetic properties of transition metal infinite 
linear chains of Fe, Co, Ni, have been calculated \cite{nau04,Spisak2,Jisang,Matej,Ederer}. 
These calculations 
show that the metallic and magnetic nanowires may become important for 
electronic/optoelectronic devices, quantum devices, magnetic storage, nanoprobes 
and spintronics. 

Despite of the above mentioned intensive theoretical and experimental research,
current understanding on novel magentic properties of nanowires and
how magnetism affects their electronic and structural properties is
still incomplete. The purpose of the present work is to make a systematic
{\it ab initio} study of the magnetic, electronic and structural properties
of both linear and zigzag atomic chains (Fig. 1) of all 3$d$ transition metals (TM).
Transition metals, because of their partly filled $d$ orbitals, have a
strong tendency to magnetize. Nonetheless, only 3$d$ transition metals
(Cr, Mn, Fe, Co, and Ni) exhibit magnetism in their bulk structures.
It is, therefore, of interest to investigate possible ferromagnetic (FM)
and antiferromagnetic (AF) magnetization in the linear chains of all 3$d$ transition
metals including Sc and Ti which appear not to have been considered.
As mentioned before, recent {\it ab initio} calculations indicate that
the zigzag chain structure of, at least, Ti\cite{li05} and Fe\cite{Spisak2} 
is energetically more favorable than the linear chain structure. 
Thus, we also study the structural, electronic and magnetic properties of 
all 3$d$ transition metal zigzag chains in order to understand how the
physical properties of the monoatomic chains evolve as their structures change
from the linear to zigzag chain. 

Relativistic electron spin-orbit coupling
(SOC) is the fundamental cause of the orbital magnetization and also the 
magnetocrystalline anisotropy energy (MAE) of solids.
The MAE of a magnetic solid is the difference in total electronic energy between 
two magnetization directions, or the energy required to rotate the magnetization
from one direction to another. It determines whether a magnet is a hard or soft one. 
Furthermore, knowledge of the MAE of nanowires is a key factor that would determine whether the nanowires
have potential applications in, e.g., high-density recording and 
magnetic memory devices. {\it Ab initio} calculations of the MAE 
have been performed for mainly the Fe and Co linear 
chains\cite{Jisang,Mokrousov2,Jisang2,aut06}, while semiempirical tight-binding calculations 
have been reported for both linear chains and two-leg
ladders of Fe and Co\cite{Druzinic,Dorantes-Davila,aut06}. 
Unlike 4$d$ and 5$d$ transition metals,
the SOC is weak in 3$d$ transition metals. Nonetheless, the MAE
could be very large in certain special 3$d$ transtion metal
structures such as tetragonal FeCo alloys\cite{bur04}. Therefore, as an endeavor
to find nanowires with a large MAE, we have calculated
the MAE and also the
magnetic dipolar (shape) anisotropy energy for all 3$d$ transition metals
in both the linear and zigzag structures. Indeed, we find that the FM
Ni linear chain has a gigantic MAE, as
will be reported in Sec. V.
Although in this paper we study only free-standing 3$d$ transition metal
chains, the underlying physical trends found may also hold for monoatomic
nanowires created transiently in break junctions\cite{Rodrigues} or encapsulated
inside 1D nanotubes\cite{Suh,Mokrousov2} or deposited on weakly interacting 
substrates~\cite{hei04}, {\it albeit}, with the actual values of the 
physical quantities being modified. 

\begin{figure}
\includegraphics[width=5cm]{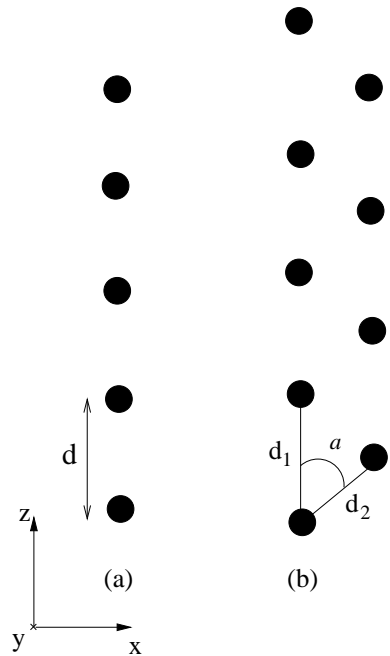}
\caption{Schematic structure diagram for (a) the linear and (b) zigzag atomic chains.}
\end{figure}

\section{Theory and Computational Method}

In the present first principles calculations, we use the accurate frozen-core full-potential 
projector augmented-wave (PAW) method,~\cite{blo94} as 
implemented in the Vienna {\it ab initio} simulation package (VASP) \cite{vasp1,vasp2}. 
The calculations are based on density functional theory with the exchange and correlation 
effects being described by the generalized gradient 
approximation (GGA)\cite{PW91}.
We adopt the standard supercell approach to model an isolated atomic chain, i.e., 
a free-standing atomic chain is simulated by a two-dimensional
array of infinite long, straight or zigzag atomic wires. 
For both linear and zigzag chains, the nearest wire-wire distance between 
the neighboring chains is, at least, 10 \AA.
This wire-wire separation should be wide enough to decouple the neighboring wires,
since the energy bands, density of states, magnetic moments and MAE from our 
test calculations with a larger separation of 15 \AA$ $ for the linear 
Fe atomic chain are nearly identical to that obtained with the wire-wire distance
of 10 \AA. A large plane-wave cutoff energy of $\sim$340 eV 
is used for all 3$d$ transition metal chains.

The equilibrium bond length (lattice constant) of the linear atomic chains in the nonmagnetic (NM),
FM and AF states is 
determined by locating the minimum in the calculated total energy as a function of the
interatomic distance. The results are also compared with that obtained by structural
optimizations, and the differences are small (within 0.4 \%) for, e.g., the
Mn, Fe and Ni chains. For the zigzag chains,
the theoretical atomic structure is determined by structural relaxations using the conjugate
gradient method. The equilibrium structure is obtained when all the forces acting on the atoms
and the axial stress are less than 0.02 eV/\AA$ $ and 2.0 kBar, respectively.   
The $\Gamma$-centered Monkhorst-Pack scheme with a $k$-mesh 
of $1\times1\times n$ ($ n = 20$) in the full Brillouin zone (BZ), 
in conjunction with the Fermi-Dirac-smearing method with $\sigma = 0.01$ eV, 
is used to generate $k$-points for the BZ integration. 
With this $k$-point mesh, the total energy is found to converge to within 10$^{-3}$ eV. 

Because of its smallness, {\it ab initio} calculation of the MAE
is computationally very demanding and needs to be carefully carried out (see, e.g.,
Refs. \onlinecite{Daalderop2,guo91}). Here we use the total energy difference approach rather
than the widely used force theorem to determine the MAE, i.e., the MAE is calculated as the
difference in the full self-consistant total energies for the two different magnetization
directions (e.g., parallel and perpendicular to the chain) concerned. The total energy
convergence criteria is 10$^{-6}$ eV/atom. A very fine $k$-point
mesh with $\sigma = 0.001$ eV is used, with $n$ being 500 for the 
linear atomic chains and 800 for the zigzag chains.
The same $k$-point mesh is used for the band structure and density of states calculations.

\section{Linear atomic chains}

\begin{figure}
\includegraphics[width=8cm]{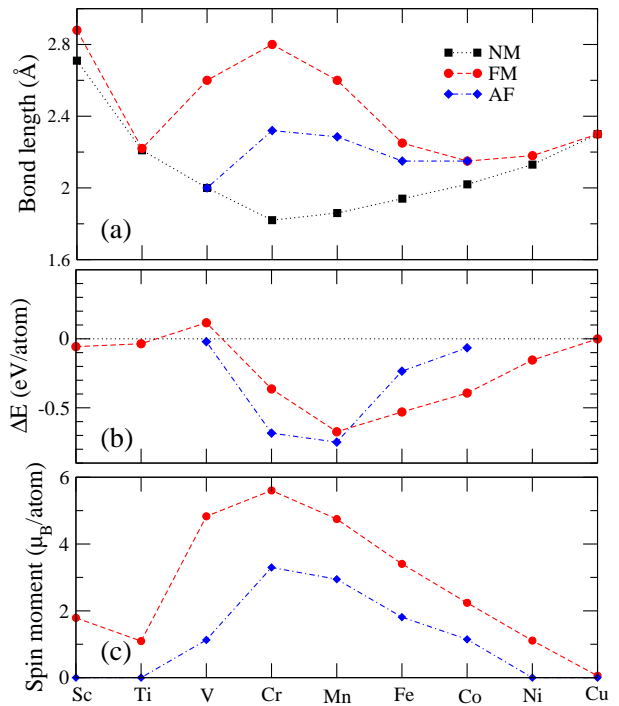}
\caption{(color online) (a) Equilibrium bond lengths, (b) magnetization energy
($\Delta E$) (i.e., the total energy of a magnetic state relative to that of
nonmagnetic state) ($\Delta E = E^{FM(AF)} - E^{NM}$) and (c) 
spin magnetic moments ($\mu_B$) of all the 3$d$ transition metal linear atomic chains
in the NM, FM, and AF states.} 
\end{figure}

\subsection{Bond length and spin magnetic moment}

The calculated equilibrium bond lengths ($d$) and atomic spin magnetic moments 
of all the 3$d$ transition metal linear chains in the NM, FM and AF states 
are displayed in Fig. 2 and also listed in Table I. 
The calculated total energy relative to that of the NM state
(i.e., the magnetization energy) of the FM and AF linear atomic chains 
are also shown in Fig. 2 and Table I. 
It is clear from Fig. 2 that 
all the 3$d$ TM elements which are nonmagnetic in their bulk structures,
become magnetic in the linear chain structures, though the FM Cu has a very small
magnetic moment and is almost degenerate with the NM state (Table I).
Furthermore, for all the 3$d$ TM elements,
the NM state is metastable and the ground state is either FM and AF 
(see Fig. 2 and Table I). The V linear chain appears to be unique in that it 
has a large FM magnetic moment of 4.8 $\mu_B$/atom but its FM state is  
higher in energy than the NM state (Table I and Fig. 2b).
Interestingly, Fig. 2 also shows that in all the cases, the equilibrium bond length is larger
in a magnetic state than in the NM state. This is due to the larger kinetic energy
in a magnetic state which make magnetic materials softer and larger in size. 
This magnetism induced increase in the bond length (or magneto-lattice expansion)
can be as large as 54 \%, as in
the case of the Cr linear chain. The ground states for the Sc, Ti, Fe, Co, and Ni chains 
are ferromagnetic while that for the V, Cr and Mn chains are antiferromagnetic. 

\begin{table}
\caption{Calculated equilibrium bond lengths ($d$) (in \AA), total energies ($E_t$)
(in eV/atom) in the FM and AF states (relative to the NM state), and spin magnetic
moments ($m_s$) (in $\mu_B$/atom), of the 3$d$ transition metal linear chains.}
\begin{ruledtabular}
\begin{tabular}{cccccccc}
   &$d_{NM}$ &$E_t^{FM}$ &$m_s^{FM}$ &$d_{FM}$ &$E_t^{AF}$ &$m_s^{AF}$ & $d_{AF}$ \\ \hline
Sc &  2.71   & -0.057    & 1.79      & 2.88    &      -    &    -      &  -    \\
Ti &  2.21   & -0.036    & 0.77      & 2.22    &      -    &    -      &  -    \\
V  &  2.00   &  0.116    & 4.06      & 2.60    &   -0.021  & 1.13      &  2.05 \\
Cr &  1.82   & -0.363    & 5.60      & 2.80    &   -0.683  & 3.30      &  2.32 \\
Mn &  1.86   & -0.673    & 4.75      & 2.60    &   -0.748  & 2.95      &  2.29 \\
Fe &  1.94   & -0.530    & 3.41      & 2.25    &   -0.235  & 1.82      &  2.15 \\
Co &  2.02   & -0.393    & 2.24      & 2.15    &   -0.066  & 1.15      &  2.15 \\
Ni &  2.13   & -0.153    & 1.11      & 2.18    &   -       &    -      &  -    \\
Cu &  2.30   & -0.001    & 0.06      & 2.30    &   -       &    -      &  -    \\
\end{tabular}
\end{ruledtabular}
\end{table}
 
The ground state bond lengths for the Sc, Ti, V, Cr, Mn, Fe, Co, Ni and Cu chains
are 2.88, 2.22, 2.05, 2.32, 2.29, 2.25, 2.15, 2.18, 2.30 \AA$ $, respectively.
These bond lengths are generally shorter than their counter-parts in the bulk structures.
For example, the calculated bond lengths for AF bcc Cr, FM bcc Fe, FM fcc Co and 
FM fcc Ni are 2.43, 2.45, 2.48 and 2.49  \AA$ $, respectively. 
The chemical bonding environment in a wire is not the same as that in a
bulk structure. In particular, the coordination number in a linear wire is certainly lower
than in a bulk material, and this may result in a shorter bond length. 
Our predictions of the FM ground state for the Fe, Co, Ni and Cu are
in good agreements with Refs. \onlinecite{nau04,Spisak2,Bahn,Nautiyal2}.
Our calculated bond lengths of the 3{\it d} TM linear chains 
in the FM state (Table I) agree well with many previous calculations. 
For example, previous
theoretical bond lengths reported for the Fe, Ni and Co chains are 2.28 \AA 
\cite{nau04}, and 2.25 \AA \cite{Spisak2} (Fe); 2.18 \AA \cite{nau04} (Co); 
2.18 \AA \cite{nau04}, and 2.16 \AA \cite{Bahn} (Ni); 2.33 \AA \cite{Bahn}, and 
2.29 \AA \cite{Nautiyal2} (Cu). Our predictions of the AF ground state for the
Cr and Mn linear chains are also consistent with the previous reports~\cite{mok07}. 
Nonetheless, the energy difference between the FM and AF states in the Fe chain
being 0.29 eV/atom, is somewhat smaller than previous
results.\cite{Jisang2}
  

\begin{figure}
\includegraphics[width=8cm]{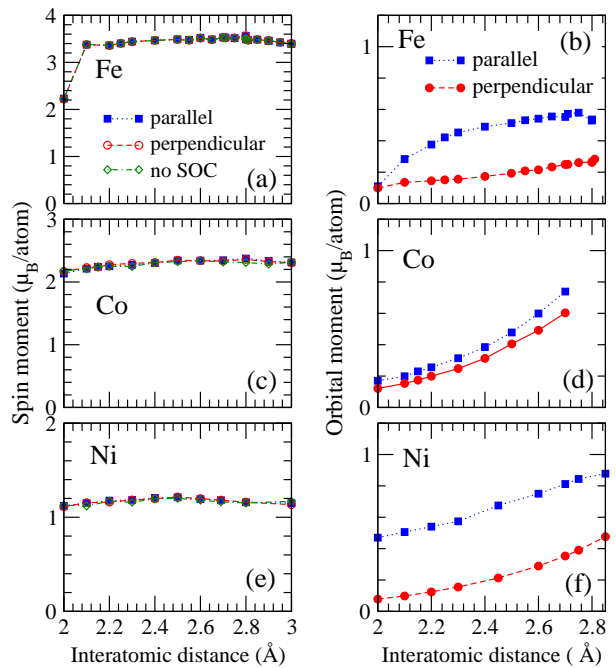}
\caption{(color online) Spin (left panels) and orbital (right panels) magnetic moments
as a function of interatomic distance of the
ferromagnetic Fe, Co, and Ni linear atomic chains. "Parallel" and "perpendicular"
denote the results from fully relativistic 
calculations with the magnetization parallel and perpendicular to the chain
axis, respectively. In the left panels, "no SOC" denote the results of the
scalar relativistic calculations.}
\end{figure}
  
\subsection{Spin-orbit coupling and orbital magnetic moment}

The relativistic SOC is essential for the orbital magnetization and
magnetocrystalline anisotropy in solids, though
it may be weak in the 3$d$ transition metal systems. Therefore, unlike several
previous studies of the magnetic properties of the 3$d$ TM 
chains,~\cite{nau04,Spisak2,mok07} we include   
the SOC in our self-consistent calculations.
When the SOC is taken into account, the spin moments for the linear atomic chains 
become 1.79 $\mu_B$ (Sc), 0.76 $\mu_B$ (Ti), 4.06 $\mu_B$ (V), 5.65 $\mu_B$ (Cr), 4.74 
$\mu_B$ (Mn), 3.41 $\mu_B$ (Fe), 2.21 $\mu_B$ (Co), 1.17 $\mu_B$ (Ni), 
0.08 $\mu_B$ (Cu), respectively. These values are almost identical to the corresponding
ones obtained without the SOC (see also Fig. 3). This is due to the weakness of the SOC in
the 3$d$ transition metals. However, including the SOC does give rise to a significant
orbital magnetic moment in some atomic chains and, importantly, allows us 
to determine the easy magnetization axis in these 3$d$ atomic chains.  
For the magnetization along the chain direction, the calculated orbital magnetic moments
in the FM state are 0.42, 0.23 and 0.45 $\mu_B$/atom for the Fe, Co and Ni chains, respectively, 
though they are only -0.04, -0.02, -0.16, -0.02, 0.04, and 0.0 $\mu_B$/atom for the Sc, Ti, V, Cr, 
Mn and Cu chains, respectively. The orbital moments in the Fe, Co and Ni atomic chains are, 
therefore, considerably enhanced, when compared with the bulk materials~\cite{guo97},
and are also larger than the orbital moments in the Fe, Co, and Ni monolayers~\cite{Guo,Guo2} 
  
To see how the magnetic properties of the atomic chain evolve with the interatomic
distance, we plot the spin and orbital moments for the Fe, Co, and Ni chains 
in the FM state as a function of the bond length in Fig. 3.
For all three 3$d$ TM chains, the spin moment remains almost unchanged as the bond 
length is increased (Fig. 3). As mentioned before, the spin moment is also unaffected when
the SOC is taken into account, due to the weakness of the SOC in 3$d$ transition metals. 
The same result is found even in the 4$d$~\cite{Mokrousov2, Delin2, Delin3, Simone}
and 5$d$~\cite{Delin} TM linear atomic chains.
Nonetheless, the SOC gives rise to rather pronounced orbital
moments in all three cases, and these orbital moments increases significantly
with the bond length, as can be seen in Fig. 3. Significantly, the orbital moment
shows a strong dependence on the magnetization orientation (Fig. 3). 
The orbital moment for the magnetization along the chain is higher than 
that for the magnetization perpendicular to the chain.
The orbital moments of the Fe, Co, 
and Ni chains at its equilibrium bond length with a perpendicular magnetization 
are only 0.15, 0.17, and 0.12 $\mu_B$, respectively.
This anisotropy in the orbital moment is especially pronounced in the
Ni chain. It is well known that in general, the magnetization direction with 
a larger orbital moment, would be lower in total energy. Therefore, the easy
magnetization direction in the Fe, Co, and Ni chains is expected to be along the chain,
as will be reported in Sec. V.
Finally, we notice that our results for the spin and orbital moments
are in good agreement with previous calculations for the Fe\cite{Mokrousov2,aut06} 
and Co\cite{Lazarovits} linear chains. 


\subsection{Band structure and density of states}
\begin{figure}
\includegraphics[width=8cm]{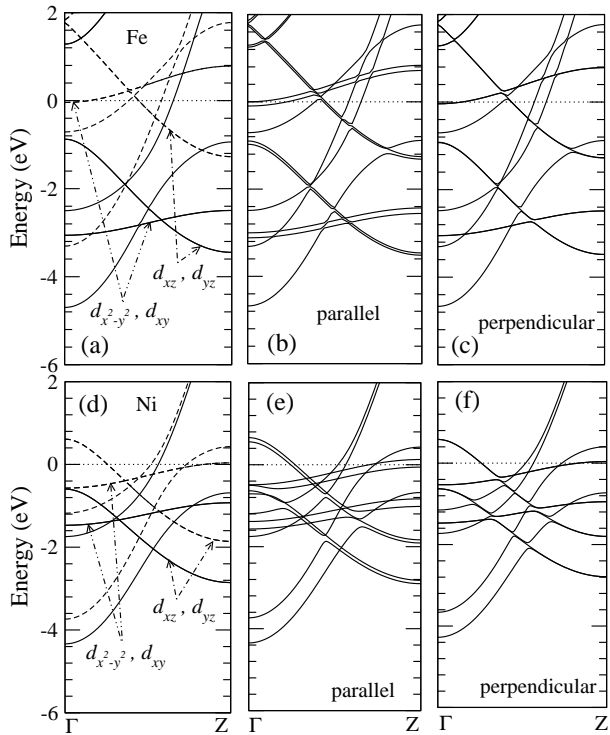}
\caption{Band structures of the Fe (upper panels) and Ni (lower panels) linear chains at 
the equilibrium bond length. Left panels: the scalar-relativistic band structures;
the middle and right panels: the fully relativistic band structures with the 
the magnetization parallel to and perpendicular to the chain axis, respectively.
In the left panels, the solid and dashed lines represent (spin up)
and (spin down) bands, respectively. The Fermi level (the dotted horizontal line)
is at the zero energy.}
\end{figure}
  
In order to understand the calculated magnetic properties, let us now examine
the band structure of the 3$d$ transition metal linear chains.
Ploted in Fig. 4 is the band structure obtained without and also with the SOC for
the Fe and Ni linear chains in the FM state at the equilibrium interatomic distance.
In the absence of the SOC, because of the linear chain symmetry, the bands
may be grouped into three sets, namely, the nondegenerate $s$- and $d_{z^2}$-dominant
bands, double degenerate ({\it d$_{ xz}$, d$_{yz}$}), and ({\it d$_{x^2-y^2}$, d$_{xy}$})
dominant bands (see the left panels in Fig. 4). The ({\it d$_{x^2-y^2}$, d$_{xy}$}) 
bands are narrow because the 
$d_{x^2-y^2}$ and $d_{xy}$ orbitals are perpendicular to the chain, thus forming
weak $\delta$ bonds. The ({\it d$_{xz}$, d$_{yz}$}) bands, on the other hands, 
are more dispersive due to the stronger overlap of the $d_{xz}$ and  $d_{yz}$ orbitals
along the chain, which gives rise to the $\pi$ bonds. The $s$- and $d_{z^2}$ dominant
bands are most dispersive since these orbitals form strong $\sigma$ bonds along the chain.
In the FM state, these bands are exchange split, and this splitting
into the spin-up and spin-down bands is 0.64, 2.58, 3.44, 3.67, 2.99, 1.94, 0.96, 
and 0.04 eV 
for the Sc, Ti, V, Cr, Mn, Fe, Co, Ni and Cu chains. The size of this spin-splitting
could be correlated with the spin moment in the FM state (see Table I). 
Also, Fig. 4 shows that when the band filling increases, as one moves from Fe to Ni, 
the ({\it d$_{x^2-y^2}$, d$_{xy}$}) bands which are partially occupied in the Fe chain, 
now lie completely below the Fermi level in the Ni chain, and hence play no role in magnetism. 
    
The directional dependence of the magnetization can be explained by analyzing
the fully relativistic band structures (see Fig. 4). For the Fe linear chain with 
the axial magnetization (Fig. 4b), the doubly degenerate {\it d$_{x^2-y^2}$, d$_{xy}$} bands 
are split into two with angular momenta {\it $m_l$= $\pm$}2. 
If one of them is fully occupied and the other is empty, the resulting orbital moment is 2. 
Nonetheless, in the Fe linear chain, both are partially occupied with different occupation numbers 
(Fig. 4b), resulting in an orbital moment of 0.42 $\mu_B$/atom. Of course,
the larger the SOC splitting, the larger the difference in the occupation number and
hence the larger the orbital moment. However, for the perpendicular magnetization,
the {\it d$_{x^2-y^2}$, d$_{xy}$} bands remain degenerate (Fig. 4c) and hence do not contribute
to orbital magnetization. Therefore, the Fe linear chain would have a smaller
orbital magnetic moment.
Of course, when the SOC is included, the degenerate {\it d$_{xz}$, d$_{yz}$} 
bands are also split into the $m_l= -1$ and +1 bands for the axial magnetization, but
remain degenerate for the perpendicular magnetization (see Fig. 4). 
This SOC splitting
of the ($d_{xz}$, $d_{yz}$) band and ($d_{x^2-y^2}$, $d_{xy}$) band is 
proportional to $|<d_{xz}|H_{SO}|d_{yz}>|^2$ and $|<d_{x^2-y^2}|H_{SO}|d_{xy}>|^2$, 
respectively. Here $H_{SO}$ is the SOC Hamiltonian. Since 
$|<d_{xz}|H_{SO}|d_{yz}>|^2$:$|<d_{x^2-y^2}|H_{SO}|d_{xy}>|^2$ = 1:4,~\cite{tak76} 
the SOC splitting of the ($d_{xz}$, $d_{yz}$) bands is much smaller than 
the ($d_{x^2-y^2}$, $d_{xy}$) bands (see Figs. 4-5). 
Therefore, the ($d_{xz}$, $d_{yz}$) bands would make a much smaller contribution
to the orbital magnetization and also the magnetocrystalline anisotropy which
will be discussed in Sec. V.

\begin{figure}
\includegraphics[width=8cm]{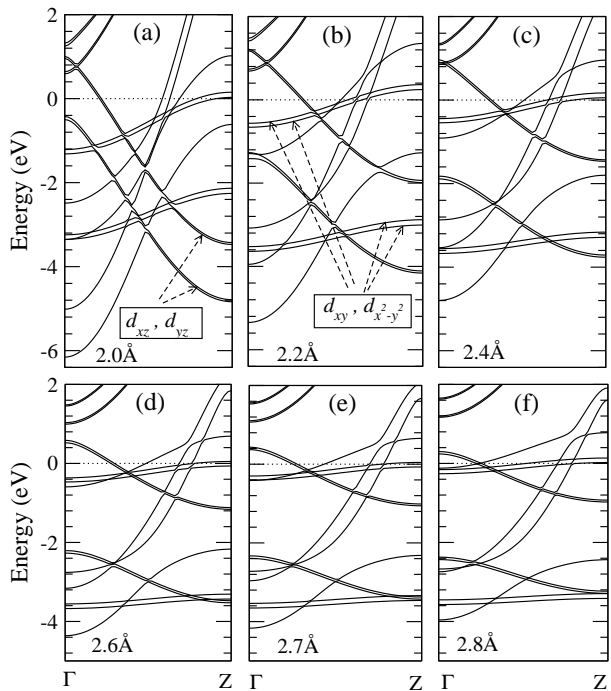}
\caption{Band structures of the Fe linear chain with the magnetization along
the chain axis at several interatomic distances. The Fermi level (the dotted horizontal
line) is the zero energy.}
\end{figure}

In Fig. 5, the band structure of the Fe linear chain at several different interatomic
distances with the magnetization along the chain direction is displayed.
Fig. 3b shows that the Fe orbital moment for both magnetization orientations
increases with the interatomic distance. However, the increase for the magnetization
along the chain axis is much more dramatic than for the perpendicular magnetization.
The reason for this variation of the orbital moment with the interatomic distance
is two fold. One is due to the increase in the localization of the 3$d$ orbital wave function
with the interatomic distance. The other is due to the detailed change in the band structure.
For example, as the interatomic distance goes from 2.4 \AA $ $ (Fig. 5c) to 2.6 \AA $ $
(Fig. 5d), one of the SOC split ({\it d$_{x^2-y^2}$, d$_{xy}$}) bands becomes fully
occupied, resulting in the increase in the occupation number difference in the 
two {\it $m_l$= $\pm$}2 bands and hence in a larger orbital moment.  

\begin{figure}
\includegraphics[width=8cm]{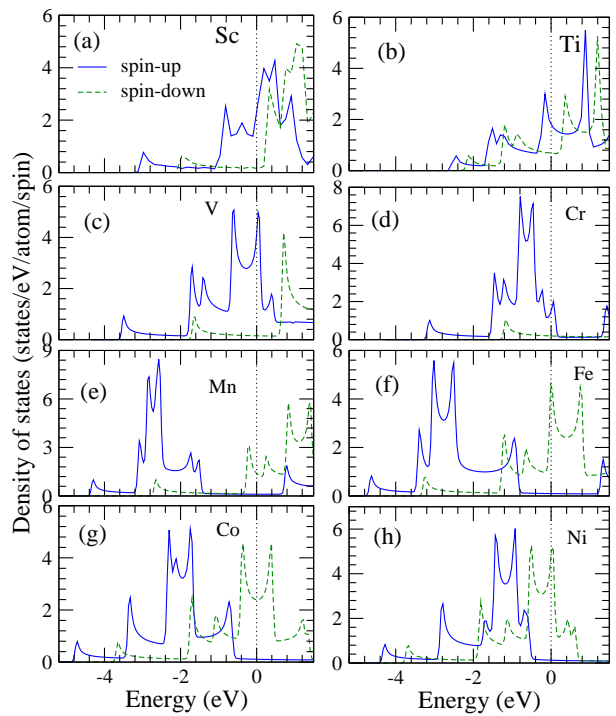}
\caption{(color online) Density of states of the FM 3{\it d} TM linear atomic chains at the equilibrium
bond length. The Fermi level (dotted vertical lines) is at the zero energy.}
\end{figure}

Electric and spin current transports are determined by the characteristics of the 
band structure near the Fermi level ($E_F$) in the systems concerned. Therefore, it would
be interesting to examine the energy bands and density of states (DOS) of the atomic
chains in the vicinity of the $E_F$. 
The spin-decomposed DOS for the FM Sc, V, Cr, Mn, Fe, Co, Ni linear chains 
are displayed in Fig. 6. In the Fe, Co, Ni cases, the spin-up states are 
nearly completely filled, and hence, the DOS at the $E_F$ is low.
On the other hand, the spin-down states are only partially occupied, resulting
in a large DOS at the $E_F$. Therefore, the density of states
at the $E_F$ in these systems are highly spin-polarized. This is usually 
quantified by the spin-polarization $P$ defined as 
\begin{equation}
P=\frac{N_{\uparrow}(E_F)-N_{\downarrow}(E_F)}{N_{\uparrow}(E_F)+N_{\downarrow}(E_F)},
\end{equation}
where $N_{\uparrow}(E_F)$ and $N_{\downarrow}(E_F)$ are the spin-up and spin-down DOS 
at the $E_F$, respectively.
The most useful materials for the spintronic applications are the so-called half-metallic
materials in which one spin channel is metallic and the other spin channel is insulating.
The spin-polarization for these half-metals is either 1.0 or -1.0, and the electric conduction
would be fully spin-polarized. 
The calculated spin-polarization and also the numbers of the conduction bands that cross the
Fermi level in the 3$d$ TM chains are listed in Table II.
It is clear that the FM V, Fe, Co, Ni linear chains have a very high spin-polarization,
though none of the 3$d$ TM linear chains in the FM state is half-metallic.
Our calculated spin-polarization for the bulk FM bcc Fe, fcc Co, and fcc Ni is -0.55, -0.77, -0.81, 
respectively, being considerably smaller than the spin-polarization of the corresponding linear chains.
This suggests that the Sc, V, Mn, Fe, Co, and Ni nanowires may be good potential materials
for spintronic devices.~\cite{wol01} Interestingly, the FM V chain has a large positive 
spin polarization (Table II and Fig. 6).

\begin{table}
\caption{Numbers ($n_c^\uparrow$ and $n_c^\downarrow$)
of the spin-up and spin-down conduction bands crossing the Fermi level, and
 spin-polarization $P$ at the Fermi level
in the 3$d$ TM atomic chains in the FM state. The spin-polarization $P$
for the bulk FM bcc Fe, fcc Co and fcc Ni are -0.55, -0.77 and -0.81, respectively}
\begin{ruledtabular}
\begin{tabular}{ccccc}
 & \multicolumn{2}{c}{linear chain}&\multicolumn{2}{c}{zigzag chain } \\
  &($n^{\uparrow}_{c}$, $n^{\downarrow}_{c}$) & $P$ &($n^{\uparrow}_{c}$, $n^{\downarrow}_{c}$) & $P$ \\ \hline
Sc     & (4, 1)  &  0.881             &(3, 0) &  0.610              \\
Ti     & (5, 4)  &  0.416             &(4, 4) &  0.181              \\
V      & (6, 1)  &  0.930             &(3, 3) &  0.085              \\
Cr     & (3, 1)  &  0.777             &(4, 2) &  0.481              \\
Mn     & (1, 3)  & -0.869             &(2, 1) & -0.829              \\
Fe     & (1, 6)  & -0.961             &(2, 4) & -0.643              \\
Co     & (1, 6)  & -0.920             &(2, 7) & -0.884              \\
Ni     & (1, 6)  & -0.951             &(2, 2) & -0.890              \\
\end{tabular}
\end{ruledtabular}
\end{table}

The magnetic properties and spin-polarization near the Fermi level in the linear
Fe, Co and Ni chains have been calculated from {\it ab initio} before by several
groups.\cite{Jisang,nau04,aut06} In particular, the free-standing FM Fe, Co and Ni linear chains
are reported to be nearly half-metallic in Ref. \onlinecite{nau04}, being in
agreement with our findings (Table II). Experimentally, a clear peak at 
0.5 $G_{0}$ ($G_{0} = 2e^2/h$ being the conductance quantum) was observed recently
in the conductance of the Co chain\cite{Rodrigues}, indicating a fully polarized conduction
in this monoatomic chain system.  

\section{Zigzag chains}
The zigzag structure for monoatomic wires has already been observed in experiments~\cite{whi91}, 
and also proposed in theoretical calculations\cite{Spisak2,Ribeiro,li05,Portal1}. 
However, among 3$d$ transition metals, only the
Ti and Fe zigzag chains have been studied theoretically~\cite{Spisak2,li05}.
In the present paper, we perform a systematic {\it ab initio} study of the structural, electronic
and magnetic properties of the zigzag chain structure of all the 3$d$ transition metals. 

\begin{table*}
\caption{Equilibrium structural parameters (see Fig. 1b for symbols $d_1$, $d_2$, $\alpha$),
spin magnetic moment ($m_s$) and magnetization energy ($\Delta E$) (i.e., the total energy
of a magnetic state relative to that of the NM state) of the 3$d$ transition metal zigzag 
chains. $d_1$ and $d_2$ are in the unit of \AA, and $\alpha$ is in the unit of degree.
$\Delta E$ is in the unit of eV/atom, and $m_s$ in the unit of $\mu_B$/atom.}
\begin{ruledtabular}
\begin{tabular}{ccccccccccc}
State&        & Sc    &   Ti    &    V  &     Cr &     Mn&     Fe&    Co &   Ni  &  Cu    \\ \hline
     &$d_1$   & 2.911 & 2.718 & 2.627 & 2.213  & 2.264 & 2.271 & 2.258 &  2.067&  2.371 \\
NM   &$d_2$   & 2.910 & 2.364 & 2.067 & 2.162  & 2.079 & 2.095 & 2.228 &  3.275&  2.334 \\
     &$\alpha$& 60.01 & 54.91 & 50.55 & 59.23  & 57.02 & 57.18 &59.56  & 71.60 & 59.56  \\ \hline
     &$d_1$   & 2.910 &  2.720&  2.626&   2.831&  2.700&  2.429& 2.318 &  2.288&  - \\
     &$d_2$   & 2.938 &  2.366&  2.066&   2.766&  2.480&  2.240& 2.218 &  2.275&  - \\
FM   &$\alpha$& 59.68 & 54.91 & 50.55 &  59.23 & 57.02 & 57.18 &58.50  & 59.82 &  - \\
     &$m_s$   & 1.050 &  0.571&  0.027&   5.904&  4.368&  3.006& 2.049 &  0.867&  - \\
  &$\Delta E$ &-0.527 & 0.002 &  0.000&  0.049 &-0.574 &-0.779 &-0.485 &-0.714 &  - \\ \hline
     &$d_1$   & - &   -   &  -    &   2.587&  2.628&  2.179& 2.313 &  2.262&  - \\
     &$d_2$   & - &   -   &  -    &   2.531&  2.413&  3.765& 2.213 &  2.301&  - \\
AF  &$\alpha$&  - &   -   &  -    &  59.27 & 57.00 & 73.18 &58.50  & 60.55 &  - \\
     &$m_s$   & - &   -   &  -    &   3.478&  3.208&  3.062& 1.311 &  0.563&  - \\
  &$\Delta E$ & - &   -   &  -    &-0.336  &-0.565 & 0.098 &-0.139 &-0.647 &  - \\

\end{tabular}
\end{ruledtabular}
\end{table*}

\subsection{ Structure and magnetic moment}
The calculated equilibrium structural parameters (Fig. 1), spin magnetic moment and magnetization energy 
of the 3$d$ TM zigzag chains 
are listed in Table III. First of all, the bond length between two nearest ions ($d_2$) in the
zigzag chains is generally similar to that ($d$) of the corresponding linear chains,
though the distance ($d_1$) between two ions in the zigzag chains along the chain direction ({\it z} axis) 
is somewhat larger than the linear chains (see Table III). 
Interestingly, most of the zigzag chains are like    
planar equilateral triangle ribbens (Fig. 1b) except the AF Fe and NM Ni zigzag chains
which look more like a sheared two-leg ladder ($\alpha \geq 71^{\circ}$, $d_1 < d_2$; see Table III). 
Note that the calculated structural parameters ($d_1$, $d_2$, $\alpha$ in
Table III) of the NM Ti zigzag chain are similar to 
that ($d_1 = 2.58$ \AA, $d_2 = 2.41$ \AA, $\alpha = 57.6$)
reported previously in Ref. \onlinecite{li05}. For a FM state of the Fe zigzag chain, 
$\alpha = 56$ and $m_s = 2.9 \mu_B$ were reported in Ref. \onlinecite{Spisak2}, being
close to the corresponding values listed in Table III here.

Secondly, all the 3$d$ TM zigzag chains except that of Cu, have magnetic solutions.
Furthermore, one can see that the Sc, Mn, Fe, Co, and Ni zigzag chains are  
most stable in the FM state, whilst the ground state of the Ti and V chains
is the NM state and that of the Cr chain is the AF state.
Note that the ground state of the linear Ti (V, Mn) chain is the FM (AF) state (Table I).
Thirdly, the spin magnetic moments in the zigzag chains (Table III) are generally 
smaller than in the corresponding linear chains (Table I). 
This is due to the increase in the coordination
number in the zigzag chains because most of them form a planar equilateral triangle ribben.
Note that our calculated spin magnetic moments for the bulk FM Fe, Co and Ni are
2.19, 1.60 and 0.63 $\mu_B$/atom, respectively, being considerably smaller than
both the linear and zigzag chains. Nonetheless, as for the linear chains,
the Cr zigzag chain still has the largest spin magnetic moment (Table III).

When the SOC is taken into account, the spin magnetic moments of the zigzag chains
remain almost unchanged, as for the linear chain cases. 
The orbital magnetic moments of the FM zigzag chains with the magnetization
along the $z$-axis are -0.003 (Sc), -0.005 (Ti), 0.000 (V), 0.000 (Cr), 0.025 (Mn),
 0.087 (Fe), 0.149 (Co), and 0.096 (Ni) $\mu_B$/atom, 
being significantly smaller than that of the corresponding linear chains (see Sec. IIIb).

\begin{figure}
\includegraphics[width=8cm]{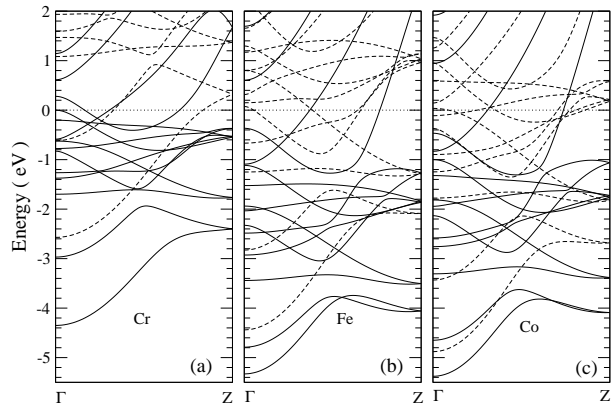}
\caption{Scalar-relativistic band structures of the Cr, Fe, 
and Co zigzag atomic chains in the FM state. 
The Fermi level (the dotted horizontal line) is at the zero energy.}
\end{figure}
\subsection{Band structures and density of states}
The scalar-relativistic band structures of the FM Cr, Fe and Co zigzag chains are
displayed in Fig. 7, as representatives. Compared with the corresponding 
band structures of the linear chains (Fig. 4a), the number of 
bands become doubled 
in the zigzag chains because of the doubling of the number of atoms. 
Furthermore, unlike the linear chains where the $d_{xy} (d_{xz})$ and
$d_{x^2-y^2} (d_{yz})$ bands (Fig. 4a and 4d) are degenerate because of rotational invariance,
the $d_{xy} (d_{xz})$ and $d_{x^2-y^2} (d_{yz})$ bands are now split because
of the strong anisotropy in the $x-y$ plane perpendicular to the chain axis.
It is clear that the energy bands are also highly spin-split
and the separation of the spin-up and spin-down bands may be correlated with the spin magnetic moment.
For example, the spin-splitting of the lowest bands is 1.78 eV
in the Cr chain, but is only 0.89 and 0.50 eV for the Fe and Co chains, respectively.


As for the linear chains, we calculate the spin-polarization ($P$) and count the numbers of spin-up
and spin-down conduction bands at the Fermi level in the FM zigzag chains, as listed in Table II.
Nevertheless, the $P$ in the V, Mn, Fe, Co and Ni zigzag chains all gets reduced (Table II).   
The largest reduction is the V zigzag chain, being reduced from 0.939 to 0.085.
The reduction is small for the Mn, Co and Ni chains (Table II), suggesting that the FM Mn, Co and
Ni zigzag chains are still useful for spintronic applications.

\subsection{Stability of linear chain structures}
Let us now compare the total energies of the linear and zigzag chains and
examine the relative stability of the two structures.
The ground state cohesive energy of the linear chains and the cohesive energies of the
zigzag chains in the NM, FM and AF states are displayed in Fig. 8.
The cohesive energy ($E_c$) of an atomic chain is defined as the 
difference between the sum of the total energy
of the free constituent atoms ($E_{a}$) and the total energy of the 
chain ($E_{t}$), i.e. $E_c = E_{a} - E_{t}$.
A positive value of the $E_c$ means that the formation of the chain from the
free atoms would save energy, i.e., the chain would be stable against breaking up
into free atoms. The total energies of the free atoms are calculated by the cubic
box supercell approach with the cell size of 10 \AA. The electronic configurations
used are  $3d^14s^2$ (Sc), $3d^34s^1$ (Ti), $3d^44s^1$ (V), $3d^54s^1$ (Cr), $3d^64s^1$ (Mn), 
$3d^74s^1$ (Fe), $3d^84s^1$ (Co), $3d^94s^1$ (Ni) and $3d^{10}4s^1$ (Cu).

\begin{figure}
\includegraphics[width=8cm]{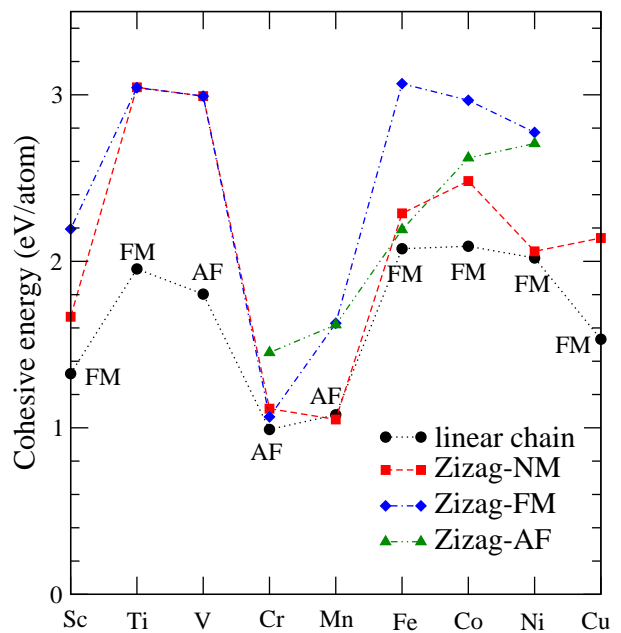}
\caption{(color online) The cohesive energy of the 3$d$ TM zigzag chains in the
NM, FM and AF states. For comparison, the ground state cohesive energy of
the corresponding linear chains is also plotted (solid circles).
The ground state magnetic configuration of the linear chains is labelled
as NM or FM or AF near each solid circle.}
\end{figure}

Remarkably, Fig. 8 shows that in all the cases, the ground state cohesive energy 
of the linear chain is smaller than that of the zigzag chain in a magnetic state. 
This suggests that the 3$d$ linear chains are unstable against the zigzag structural distortion, 
as may be expected from the Peierls instability of
linear one-dimensional monoatomic metals.~\cite{kit96}
The difference in the ground state energy between the linear and zigzag structures
for all the 3$d$ elements is rather large, ranging from 0.8 to 1.1 eV/atom. 
This shows that the free standing 3$d$ TM linear chains would not be the stable state,
and the linear chains may occur only in constrained conditions such as on the
steps on a vicinal surface~\cite{Gambardella2} and under tensile stress in 
the break-point experiments~\cite{Rodrigues,Rodrigues2,Jiandong,Susumu,Ohnishi}.
Only the zigzag structure of Fe has been considered before in Ref. ~\onlinecite{Spisak2}.
The zigzag structure of Fe was also found to be lower in energy than the
linear chain by 1.01 eV/atom, a value being close to 0.99 eV/atom found in the
present calculations. 

Interestingly, we find that for some 3$d$ transition metals, the
ground state magnetic configuration changes when the structure changes from
the linear to zigzag chain.
For example, the ground state of the Sc and Ti chains is ferromagnetic in the linear chain 
but becomes nonmagnetic in the zigzag structure (Table III).
On the other hand, the ground state of the Mn chain changes from the AF in the linear chain
to the FM state in the zigzag structure. Nevertheless, the total energy difference
between the FM and AF states in the Mn zigzag chain is small (within $\sim$0.01 eV/atom).
The ground state of the V chain also changes, from the AF state in the linear chain
to the FM state in the zigzag chain.

\section{Magnetic anisotropy energy}

The total energy as a function of the magnetization orientation ($\theta,\phi$)
of a 1D wire may be written, in the lowest non-vanishing terms, as 
\begin{equation}
E_t=E_{0}+sin^{2}\theta (E_{1}-E_{2}cos^{2}\phi)
\end{equation} 
where $\theta$ is the polar angle of the magnetization away from the chain axis ($z$-axis)
and $\phi$ is the azimuthal angle in the $x-y$ plane perpendicular to the wire, measured from the 
{\it x} axis. For the free standing linear atomic chains, the azimuthal anisotropy
energy constant $E_{2}$ is zero, because of the rotational invariance. 
The axial anisotropy energy constant $E_1$ is then given by the total energy
difference between the magnetization along the $y$($x$) and {\it z} axes,
i.e., $E_1 = E^y - E^z$ ($E^x = E^y$). A positive value of $E_1$ means that the chain ($z$) axis 
is the easy magnetization axis. For the zigzag chains which are in the $x-z$ plane,
$E_{2}$ is not zero and can be calculated as the total energy difference
between the magnetization along the $x$ and $y$ axes, i.e., $E_2 = E^y - E^x$.

\begin{table}
\caption{Total ($E_1^t$), electronic ($E_1^e$) and dipolar ($E_1^d$) magnetic anisotropy energes (in meV/atom) of the 3$d$ transition metal linear chains.
If $E_1^t$ is positive, the easy magnetization axis is along the chain; otherwise, the easy magnetization axis is perpendicular to
the chain.}
\begin{ruledtabular}
\begin{tabular}{ccccccc}
   & \multicolumn{3}{c}{FM} & \multicolumn{3}{c}{AF} \\
   & $E_1^t$ & $E_1^e$ & $E_1^d$ & $E_1^t$ & $E_1^e$ & $E_1^d$ \\ \hline
Sc & 0.06  &  0.01  &  0.05  &   - &  - &  -  \\
Ti & -0.22 & -0.27  &  0.05  &   - &  - &  -  \\
V  & 0.81  &  0.45  &  0.36  & 0.183  & 0.138 & 0.045 \\
Cr & 0.62  &  0.07  &  0.55  & -0.006 & -0.259 &  0.253  \\
Mn & 0.22  & -0.28  &  0.50  & -0.487 & -0.698 &  0.212  \\
Fe & 2.65  &  2.25  &  0.40  & -0.817 & -0.917 &  0.100  \\
Co & -0.48  & -0.68  &  0.20  &  5.194 &  5.155 &  0.039  \\
Ni &11.44  & 11.39  &  0.05  &   -  &  - &  -  \\
\end{tabular}
\end{ruledtabular}
\end{table}

The magnetic anisotropy energy for a magnetic solid consists of two contributions.
One comes from the magnetocrystalline anisotropy in the electronic band structure
caused by the simultaneous occurrence of the electron spin-orbit interaction
and spin-polarization in the magnetic system. This is known as the electronic contribution and
{\it ab initio} calculation of this part has already been described in Sec. II.
The other is the magnetostatic (or shape) anisotropy energy due to the magnetic dipolar
interaction in the solid. The shape anisotropy energy is zero for the cubic systems such as bcc Fe and fcc Ni,
and also negligibly small for weakly anisotropic solids such as hcp Co. 
However, for the highly anisotropic structures such as magnetic Fe and Co monolayers,~\cite{Guo,Guo2}
the shape anisotropy energy can be comparable to the electronic MAE, and therefore cannot be
be neglected. Furthermore, as will be reported immediately below,
the shape anisotropy energy of the 3$d$ TM atomic chains are also large and cannot be ignored.
Therefore, in this work, following Ref. \onlinecite{Guo}, we use Ewald's lattice summation
technique\cite{Ewald} to calculate the magnetic dipole-dipole interaction energy.
For the collinear magnetic systems ({\it{i.e.}} {\bf{m$_{q}$}}//{\bf{m$_{q^{'}}$}}), 
this magnetic dipolar energy {\it $E_{d}$} is given by (in atomic Rydberg units)~\cite{Guo}
\begin{equation}
 E^{d} = \sum_{qq^{'}}{\frac{m_{q}m_{q^{'}}}{c^2} M_{qq^{'}}}
\end{equation}
and
\begin{equation}
 M_{qq^{'}}= {\sum_{\bf R}}^{'}\frac{1}{\mid \bf R+q+q^{'}\mid^{3}}\{1-3\frac{[(\bf R+q+q^{'})\cdot \hat{m_q}]^2}{\mid \bf R+q+q^{'}\mid^{2}}\}
\end{equation}
where  $M_{qq^{'}}$ is called the magnetic dipolar Madelung constant which is evaluated
by Ewald's lattice summation technique \cite{Ewald}. The speed of 
light $c = 274.072$. {\bf R} are the lattice vectors, {\bf q} 
are the atomic position vectors in the unit cell and {$m_{q}$} is the atomic magnetic moment
on site {\it q}. Note that in atomic Rydberg units, one Bohr magneton ($\mu_B$) is $\sqrt{2}$.
Therefore, the magnetic dipolar energy $E^d$ for the multilayers obtained previously by Guo and 
co-workers\cite{Guo,Guo2,guo92,guo94,guo99} is too small by a factor of $2$. 

\begin{table*}
\caption{The total ($E_1^t$, $E_2^t$), electronic ($E_1^e$, $E_2^e$) and dipolar ($E_1^d$, $E_2^d$) magnetic
anisotropy energy constants (in meV/atom) as well as the easy magnetization axis
({\bf M}) of the 3$d$ transition metal zigzag chains.
$E_1$ = $E^y$ - $E^z$; $E_2$ = $E^y$ - $E^x$, see Eq. (2).}
\begin{ruledtabular}
\begin{tabular}{ccccccccccccccc}
   & \multicolumn{5}{c}{FM} &\multicolumn{5}{c}{AF}  \\
   &$E_1^e$ &$E_2^e$ &$E_1^d$ &$E_2^d$ & $E_1^t$ &$E_2^t$ & {\bf M}&$E_1^e$ &$E_2^e$ &$E_1^d$ &$E_2^d$ & $E_1^t$ &$E_2^t$ & {\bf M}\\ \hline
Sc & 0.000& 0.000&0.022&0.011&0.022&0.011&$z$ &  -   &   -  &  -   &  -   &  -  &  -  &- \\
Ti & 0.000& 0.000&0.010&0.006&0.010&0.006&$z$ &  -   &   -  &  -   &  -   &  -  &  -  &- \\
V  & 0.000& 0.000&0.000&0.000&0.000&0.000& -  &  -   &   -  &  -   &  -   &  -  &  -  &- \\
Cr &-0.158&-0.158&0.857&0.418&0.699&0.260&$z$ &-2.663&-2.754& 0.152&-0.190&-2.511&-2.944&$y$ \\
Mn &-0.194&-0.256&0.578&0.299&0.384&0.043&$z$ & 0.898& 1.024& 0.091&-0.190& 0.989&0.834&$z$ \\
Fe &-0.700&-0.584&0.374&0.193&-0.326&-0.391&$y$ & 0.012& 0.006& 0.315&-0.071&0.327 &-0.065&$z$ \\
Co & 1.317& 1.122&0.192&0.096&1.509&1.218&$z$ & 8.763& 6.992& 0.029&-0.040&8.792&6.952&$z$\\
Ni & 0.636& 1.741&0.035&0.017&0.671&1.758&$x$ & 6.911& 5.126& 0.006&-0.007&6.917&5.119&$z$\\
\end{tabular}
\end{ruledtabular}
\end{table*}

The calculated shape anisotropy energies ($E^d$) for the linear and zigzag chains 
are listed in Tables IV and V, respectively. Tables IV and V show that in both the linear 
and zigzag chains and in both the FM and AF states,
the shape anisotropy energies can be comparable to the electronic contributions.
Furthermore, they always prefer the chain direction ($z$ axis) as the easy magnetization axis,
and this may be expected since the shape anisotropy energy always favors the direction of the 
longest dimension. Therefore, any perpendicular magnetic anisotropy must originate from the 
electronic magnetocrystalline anisotropy when it overcomes the shape anisotropy.
In the zigzag chains, there is also a significant magnetic anisotropy in the
$x-y$ plane perpendicular to the chain axis. For the FM state, the $x$ axis is favored in
all the zigzag chains,
i.e., the $y$ axis would be the hard magnetization axis if the magnetic anisotropy 
were determined by the $E^d$ alone. In contrast, for the AF state, the $x$ axis 
would be the hard axis (see Table V).

The calculated electronic anisotropy energies of the linear and zigzag 
atomic chains are also listed in Tables IV and V, respectively.
Interestingly, Table IV shows that in the FM linear chains, the electronic anisotropy 
energy would favor a perpendicular anisotropy in the Ti, Mn, Co chains but prefer the 
chain axis in the Sc, V, Cr, Fe and Ni chains. Nevertheless, the easy magnetization 
direction is predicted to be the chain axis in all the 3$d$ FM linear chains except Ti 
and Co because the perpendicular electronic 
anisotropy in the Mn chain is not sufficiently large to overcome the axial 
shape anisotropy (Table IV).
In the AF state, in contrast, the Cr, Mn and Fe linear chains would have the easy axis perpendicular
to the chain while the V and Co linear chains still prefer the axial anisotropy. 
Remarkably, the FM Ni linear chain has a gigantic axial anisotropy energy (Table IV),
being in the same order of magnitude of that in the 4$d$ transition metal 
linear chains~\cite{Mokrousov}. In the 4$d$ transition metals, the SOC splittings are 
large, being about ten times larger than the 3$d$ transition metals, and thus the
large MAE in the 4$d$ transition metal linear chains
may be expected. The axial anisotropy energy for the FM V, Cr, Fe and AF Co linear
chains are also rather large, being generally a few times larger than the corresponding
monolayers.~\cite{Guo,Guo2} 

\begin{figure}
\includegraphics[width=8cm]{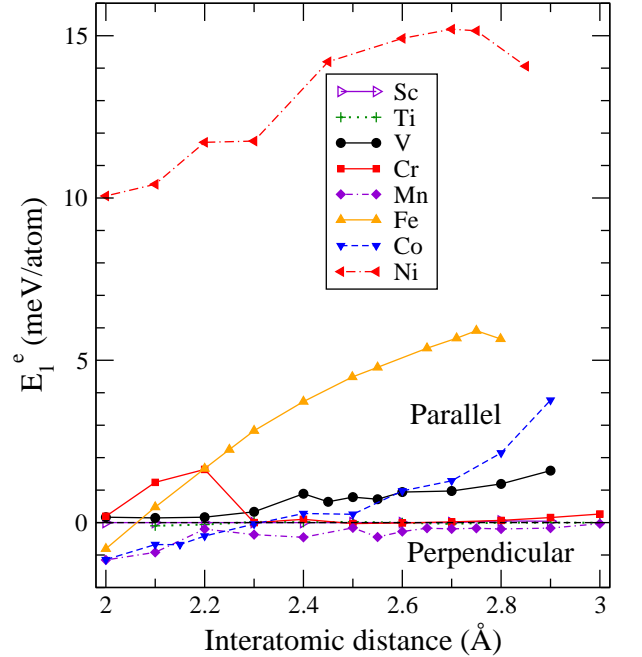}
\caption{(color online) Magnetocrystalline anisotropy energy ($E_1^e$) of the 3$d$ transition
metal linear atomic chain as a function of interatomic distance.
A positive value of $E_1^e$ means that the magnetization would be parallel to the chain axis whilst a negative value would means that the easy magnetization axis would be
perpendicular to the chain.} \end{figure}

In Fig. 9, the electronic anisotropy energy for the linear 3$d$ atomic chains is displayed as
a function of the lattice constant (bond length).
It is clear that in all the 3$d$ linear chains except that of Fe and Co, 
the easy axis of magnetization remains the same no matter the chain is elongated or 
compressed. 
When compressed, the FM Fe linear chain would undergo a spin reorientation transition
from the axial to perpendicular direction at the bondlength of $\sim$2.06\AA.
In contrast, the FM Co linear chain would transform from the perpendicular to axial
direction at $\sim$2.31 \AA$ $ when elongated.

To help identifying the electronic origin of the magnetocrystalline anisotropy, 
we plot the scalar-relativistic $d$-orbital decomposed DOS for the FM Fe, Co and
Ni linear chains in Fig. 10.  
According to perturbation theory analysis, the occupied and empty $d$-states
in the vicinity of the Fermi level which are coupled by the SOC are
most important to the magnetocrystalline anisotropy~\cite{wan93}. 
Furthermore, the SOC matrix elements 
$<d_{xz}|H_{SO}|d_{yz}>$ and $<d_{x^2-y^2}|H_{SO}|d_{xy}>$ are found to 
contribute to the axial anisotropy 
while $<d_{x^2-y^2}|H_{SO}|d_{yz}>$, $<d_{xy}|H_{SO}|d_{xz}>$ 
and $<d_{3z^2-r^2}|H_{SO}|d_{yz}>$ prefer a perpendicular anisotropy.~\cite{tak76}
The ratio of these matrix elements are $<d_{xz}|H_{SO}|d_{yz}>^2$:$<d_{x^2-y^2}|H_{SO}|d_{xy}>^2$:
$<d_{x^2-y^2}|H_{SO}|d_{yz}>^2$:$<d_{xy}|H_{SO}|d_{xz}>^2$:$<d_{3z^2-r^2}|H_{SO}|d_{yz}>|^2$
=$1:4:1:1:3$.~\cite{tak76}
Fig. 10 (a,d) shows that in the FM Fe and Ni chains, the $E_F$ sits on, respectively, 
the lower and upper sharp peak of the spin-down $d_{x^2-y^2}$ and $d_{xy}$ DOS. Consequently,
the SOC near the $E_F$ between the $d_{x^2-y^2}$ and $d_{xy}$ bands
would give rise to a dominating contribution to the magnetocrystalline anisotropy
and therefore, the FM Fe and Ni chains would prefer the chain axis.
On the other hand, in the FM Co chain, the $E_F$ lies in the valley of the 
spin-down $d_{x^2-y^2}$ and $d_{xy}$ DOS (Fig. 10 b). As a result, 
the $<d_{x^2-y^2}|H_{SO}|d_{xy}>$ contribution does not dominate and the FM Co chain
would perfer a perpendicular anisotropy. Nonetheless, the spin-down $d_{x^2-y^2}$ 
and $d_{xy}$ DOS near the $E_F$ increases dramatically when the FM Co chain is stretched
(Fig. 10 c). This would enhance the $<d_{x^2-y^2}|H_{SO}|d_{xy}>$ contribution considerably
and the FM Co chain would then prefer the axial anisotropy when the bondlength
is larger than $\sim$2.23 \AA$ $ (Fig. 9).

\begin{figure}
\includegraphics[width=8cm]{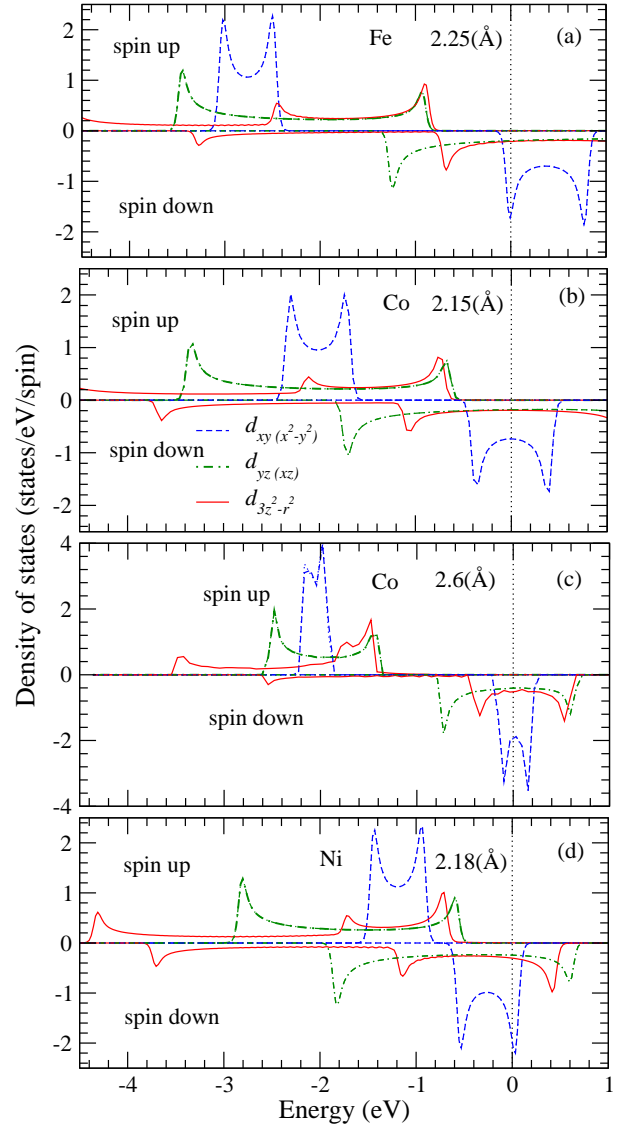}
\caption{(color online) $d$-orbital-decomposed density of states of the FM (a) Fe, (b) Co 
and (d) Ni linear chains at the equilibrium bond length. In (c), $d$-orbital-decomposed 
density of states of the FM Co linear chain at an elongated bond length of
2.6 \AA$ $ is displayed. }
\end{figure}

Generally speaking, the magnetic anisotropy in at least the FM 3$d$ chains becomes smaller
as the structure moves from the linear to zigzag structure (see Tables IV and V). 
The most dramatic reduction in the magnetic anisotropy occurs in the FM Ni chain.
The axial anisotropy constant $E_1$ in the zigzag Ni chain is one order of magnitude
smaller than that in the linear chain. There is now a large anisotropy 
energy ($E_2$) in the $x-y$ plane perpendicular to the chain axis (see Table V). 
As a result, the easy magnetization axis in the zigzag FM Ni chain is
in the zigzag chain plane but perpendicular to the chain axis, i.e., the $x$-axis
(see Fig. 1). For the FM Ti and Co chains, the easy magnetization changes from the perpendicular to
axial direction (Tables IV and V). Strikingly, no AF state
could be stabilized in the linear Ni chain, and in contrast, the AF state not only can
be stabilized but also has gigantic magnetic anisotropy energies in the zigzag chain (see Table V).  
Table V also shows that the magnetic anisotropy energies in the zigzag AF Co chain
are considerably enhanced compared with that in the linear AF Co chain.

\section{Conclusions}

We have performed a systematic {\it ab initio} study of the magnetic and 
electronic properties of both linear and zigzag atomic chains of all 3$d$ transition 
metals within density functional theory with GGA.
The accurate frozen-core
full-potential PAW method is used. The underlying atomic structures
were determined theoretically. All the zigzag chains except the NM Ni
and AF Fe chains which form a twisted two-legger ladder,
look like a corner-sharing triangle ribbon, and have a lower
total energy than the corresponding linear chains.

We find that all the 3$d$ transition metals in both linear and zigzag structures
have a stable or metastable FM state. 
Furthermore, in the V, Cr, Mn, Fe, Co linear chains
and Cr, Mn, Fe, Co, Ni zigzag chains, a stable or metastable AF state exists too.
In the Sc, Ti, Fe, Co, Ni linear structures, the FM state is the ground state whilst
in the V, Cr and Mn linear chains, the AF state is most energetically favorable.
The electronic spin-polarization at the Fermi level in the FM Sc, V, Mn, Fe,
Co and Ni linear chains is close to 90\% or above, suggesting that these nanostructures
may have applications in spin-transport devices. Only in the Cr zigzag structure, the AF state
is energetically more favorable than the FM state.
Surprisingly, the V, Cr, Mn, and Fe linear chains show a giant magneto-lattice expansion
of up to 54 \%.

Both the electronic magnetocrystalline anisotropy energy
and magnetic dipolar anisotropy energy
have been calculated. We find that shape anisotropy energy
can be comparable to the electronic one and always prefer the axial magnetization
in both the linear and zigzag structures. Furthermore,
in the zigzag chains, there is also a pronounced shape anisotropy in the
plane perpendicular to the chain axis.
Nonetheless, in the FM Ti, Mn, Co and AF Cr, Mn, Fe linear chains, 
the electronic anisotropy is perpendicular,
and it is sufficiently large in the FM Ti and Co as well as the AF Cr, Mn 
and Fe linear chains such that the
easy magnetization axis is perpendicular.
In the AF Cr and FM Ni zigzag structures, the easy magnetization direction
is also perpendicular to the chain axis but in the ribbon plane.
Remarkably, the axial magnetic anisotropy in the FM Ni linear chain is
gigantic, being $\sim$ 12 meV/atom, suggesting that Ni nanowires could have important
applications in ultrahigh density magnetic memories and hard disks.
The axial magnetic anisotropy energy of the FM V, Cr, Fe linear chains
and FM Cr, Mn, Co zigzag structures is also sizable.
Interestingly, there is a spin-reorientation transition in the FM Fe and Co linear
chains when the chains are compressed or elongated.
Large orbital magnetic moment is found in the FM Fe, Co and Ni linear chains.
Finally, the electronic band structure and density of states of the nanowires
have also been calculated mainly in order to understand the electronic origin
of the large magnetocrystalline anisotropy and orbital magnetic moment.

\section*{Acknowledgments}
The authors thank Z.-Z. Zhu for stimulating discussions on zigzag monatomic chains.
The authors acknowledge supports from National Science Council and NCTS of Taiwan. 
They also thank National Center for High-performance Computing of Taiwan 
and NTU Computer and Information Networking Center for providing CPU time.

\newpage 
\bibliography{99}

\end{document}